\documentclass[twocolumn]{aastex631}

\shorttitle{Towards a live homogeneous database of solar active regions}
\shortauthors{Ruihui Wang et al.}

\graphicspath{{./}{figures/}}
\usepackage{multirow}
\usepackage{booktabs}
\usepackage{hyperref}
\usepackage{amsmath}
\usepackage{threeparttable}

\begin{document}

\title{Towards a live homogeneous database of solar active regions based on SOHO/MDI and SDO/HMI synoptic magnetograms. I. Automatic detection and calibration}

\author{Ruihui Wang}
\affiliation{ School of Space and Environment, Beihang University, Beijing, People’s Republic of China; jiejiang@buaa.edu.cn}
\affiliation{Key Laboratory of Space Environment Monitoring and Information Processing of MIIT, Beijing, People’s Republic of China}

\author{Jie Jiang}
\affiliation{ School of Space and Environment, Beihang University, Beijing, People’s Republic of China; jiejiang@buaa.edu.cn}
\affiliation{Key Laboratory of Space Environment Monitoring and Information Processing of MIIT, Beijing, People’s Republic of China}

\author{Yukun Luo}
\affiliation{ School of Space and Environment, Beihang University, Beijing, People’s Republic of China; jiejiang@buaa.edu.cn}
\affiliation{Key Laboratory of Space Environment Monitoring and Information Processing of MIIT, Beijing, People’s Republic of China}

\begin{abstract}
Recent studies indicate that a small number of rogue solar active regions (ARs) may have a significant impact on the end-of-cycle polar field and the long-term behavior of solar activity. The impact of individual ARs can be qualified based on their magnetic field distribution. This motivates us to build a live homogeneous AR database in a series of papers. As the first of the series, we develop a method to automatically detect ARs from 1996 onwards based on SOHO/MDI and SDO/HMI synoptic magnetograms. The method shows its advantages in excluding decayed ARs and unipolar regions and being compatible with any available synoptic magnetograms. The identified AR flux and area are calibrated based on the co-temporal SDO/HMI and SOHO/MDI data. The homogeneity and reliability of the database are further verified by comparing it with other relevant databases. We find that ARs with weaker flux have a weaker cycle dependence. Stronger ARs show the weaker cycle 24 compared with cycle 23. Several basic parameters, namely, location, area, and flux of negative and positive polarities of identified ARs are provided in the paper. This paves the way for AR's new parameters quantifying the impact on the long-term behavior of solar activity to be presented in the subsequent paper of the series. The constantly updated database covering more than two full solar cycles will be beneficial for the understanding and prediction of the solar cycle. The database and the detection codes are accessible online.
\end{abstract}


\section{Introduction} \label{sec:intro}
Active regions (ARs) on the Sun are places where the strong magnetic field is distributed. They originate from the interior dynamo process and correspond to toroidal magnetic flux emergence. The properties of ARs' emergence and the subsequent decay can provide vital clues for large-scale dynamo models. The ARs also provide the seed field for subsequent cycles in the Babcock-Leighton (BL) dynamo framework \citep{Babcock1961,Leighton1969}.

In the context of the BL dynamo, the emergence and subsequent transport of tilted ARs on the solar surface account for the generation of the Sun’s poloidal field, in particular its dipole component represented by the polar fields. The poloidal field is stretched by differential rotation (quasi-)linearly to regenerate the toroidal field for the AR emergence of the subsequent cycle. Hence the correlation between the minima of the polar field and the amplitude of the next cycle is expected \citep{Schatten1978} and also confirmed by direct polar field observations \citep{Jiang2007} and polar field proxies \citep{Wang2009, Munoz-Jaramillo2013}. Hence if we can predict an AR's contribution to the end-of-cycle polar field (or axial dipole field), we can evaluate its impact on the subsequent solar activity. 

ARs are always approximated as bipolar magnetic regions (BMRs) having symmetric leading and following polarities in morphology when they are involved with the study of the solar cycle. The initial contribution of a newly-emergent BMR to the axial dipole field ($D_{\mathrm{BMR}}^{i}$) satisfies $D_{\mathrm{BMR}}^{i} \propto A^{\frac{3}{2}}\sin\alpha\cos\lambda$ \citep{Wang1991, Yeates2023}, where A, $\alpha$, and $\lambda$ are area, tilt angle, and latitude of each BMR, respectively.

But after a BMR has emerged, meridional flow and supergranular diffusion acting in combination can cause its axial dipole field to grow or decay, depending on the BMR’s emerging latitude $\lambda$ \citep{Wang1991}. The final contribution of the BMR to the dipole field $D_{\mathrm{BMR}}^{f}$ at the end of a cycle obeys
\begin{equation}
\label{eqn:dip_bmr_final}
	D_{\mathrm{BMR}}^{f}\propto D_{\mathrm{BMR}}^{i}* \exp\left(-\frac{\lambda^2}{2\lambda_R^2}\right), 
\end{equation}
where $\lambda_R$ depends on the transport processes  \citep{Jiang2014, Whitbread2018, Petrovay2020b}. This indicates that a single AR with a large flux and tilt angle emerging at low latitudes can have a dramatic impact on the dipole field at the end of the cycle and the further course of cyclic solar magnetic activity \citep{Jiang2015}. Although ARs emergence shows systematic properties in their latitude and tilt angle \citep{Jiang2011}, there are also strong stochastic components. At low latitudes, large ARs with large tilt angles can emerge randomly during a solar cycle. Hence they are referred to as rogue ARs \citep{Nagy2017}. From the point of view of solar cycle prediction, it is important to define a parameter describing the deviation of the dipole contribution from the case with no stochastic perturbations in AR emergence \citep{Petrovay2020}. The parameter was first proposed by \cite{Nagy2020} as the degree of rogueness, that is ARDoR in abbreviation. 

However, realistically ARs are not BMRs. They have various configurations \citep{Hale1919, Kunzel1960}. \cite{Jaeggli2016} indicate that about 30\% $\beta$-type ARs observed during the years of solar maxima are appended with the classifications $\gamma$ and/or $\delta$, which usually have large areas and are hard to quantify their realistic tilt angles because of the complex multipolar configuration. \cite{Jiang2019, 2020Yeates} assimilate ARs' real magnetic configuration into their surface flux transport (SFT) simulations. The final dipole field $D_{\mathrm{BMR}}^{f}$ and the initial dipole field $D_{\mathrm{BMR}}^{i}$ does not obey Eq.(\ref{eqn:dip_bmr_final}) anymore. The changing sign between the two parameters is common for $\delta$-type spots. Hence the realistic AR magnetic field distribution is required to predict individual ARs' contribution to the final dipole accurately. \cite{Wang2021} further provide a quick and precise quantification of the contribution of an AR with the central latitude $\lambda$ to the final dipole field $D_{\mathrm{AR}}^{f}$ instead of SFT simulations as
\begin{equation}
	\label{eqn:dip_AR_final}
	D_{\mathrm{AR}}^{f}\propto\iint B(\theta,\phi)\mathrm{erf}\left(|\lambda|/\sqrt{2}\lambda_R \right)\mathrm{sgn}(\lambda)\sin \theta d\theta d\phi , 
\end{equation}
where  $B(\theta,\phi)$ is the magnetic field distribution of the identified AR, $\theta$ and $\phi$ are the co-latitude and longitude, respectively. Thus with $B(\theta,\phi)$, we may predict the AR's contribution to the end-of-cycle polar field quickly. 

Although a small number of rogue ARs cause large variations of the polar field, the cumulative effect of many weaker regions is also significant \citep{Whitbread2018, Hofer2023}. Hence for a better understanding of the solar cycle, especially the variability of the polar field, we need a complete AR catalog. Moreover, to achieve this end the database is required to be long-term and homogeneous. Constant updates of the database are also required for monitoring and predicting AR's impact on solar cycles.

The aforementioned progress in understanding the effect of individual ARs on the solar cycle motivates us to develop a live homogeneous database of ARs for a better understanding and prediction of the solar cycle. So far several AR databases are available already, for example, RGO and USAF/NOAA AR Database \footnote{http://solarcyclescience.com/activeregions.html}, Bipolar Active Region Detection \citep[BARD,][]{MunozJaramillo2021}, Space-weather HMI Active Region Patches \citep[SHARPs,][]{Bobra2014}, Space-Weather MDI Active Region Patches \citep[SMARPs,][]{SMARPsASHARPs} and \cite{Sreedevi2023}. But these databases just provide ARs’ parameters relevant to space weather effect. \cite{2020Yeates} gives an exceptional database, in which both the initial and final dipole field of each AR calculated from SFT simulations are offered. The two space-based instruments, SOHO/MDI \citep{MDI} and SDO/HMI \citep{HMI}, provide continuous, seamless, and high-resolution synoptic magnetograms over solar cycles 23, 24, and 25, from 1996 to the present day. These magnetograms provide us the opportunity to build the AR database for the understanding and prediction of the solar cycle.

This paper commences a series of studies toward a live homogeneous database of solar ARs based on SOHO/MDI and SDO/HMI synoptic magnetograms. In the first paper, we develop a method compatible with any available synoptic magnetograms to automatically detect ARs. The identified AR flux is calibrated based on the co-temporal SDO/HMI and SOHO/MDI data. The homogeneity and reliability of the database are further verified by comparing it with other relevant data. Several basic AR parameters, namely, location, area, and flux of negative and positive polarities of identified ARs are provided in the paper. In the second paper, we will provide and analyze parameters quantifying the impact of individual ARs on the long-term behavior of solar activity, e.g., the final contribution to the dipole field and degree of rogueness based on the automatically identified ARs in the first paper. 

This paper is organized as follows. In Section \ref{sec:methods}, we describe the algorithms for AR automatic detection. In Section \ref{sec:calibration}, we calibrate the detected results based on the co-temporal SDO/HMI and SOHO/MDI data. In Section \ref{sec:comparison}, we compare our data with other available ones to evaluate our method and the homogeneity of the data. In Section \ref{sec:parameters}, we overview the properties of detected ARs in our database. We summarize and discuss the above results in Section \ref{sec:conclusion}.

\section{Automatic detection Method} \label{sec:methods}

\subsection{Observed synoptic magnetograms}\label{subsec:data}
The data used in this study are radial synoptic magnetograms that are constructed from full-disk magnetograms obtained by Michelson Doppler Imager on board the Solar and Heliospheric Observatory (SOHO/MDI) \citep{MDI} and Helioseismic and Magnetic Imager on board the Solar Dynamics Observatory (SDO/HMI) \citep{HMI}. 

The sizes of MDI and HMI synoptic magnetograms are both $1440\times3600$ pixels. Each pixel is 0.1 Carrington longitude and about 0.00139 sine-latitude. Its area is about 1.17 Mm$^2$. The time range of MDI and HMI magnetograms is from CR 1909 (May 1996) to CR 2096 (May 2010) and from CR 2097 (June 2010) to CR 2265 (December 2022), respectively, except for CRs 1938-1940 when the MDI maps are totally missing. Besides, the data of MDI maps are partially missing in CRs 1909, 1910, 1937, 1941-1946, 1956, 1986, 2004, 2005, 2011, 2015, 2052, and 2086.  All missing data mentioned above are due to the SOHO spacecraft malfunction. The ARs in the maps with partially missing data are still detected but need to be used with care.

\subsection{Algorithm of Automatic detection}\label{subsec:alg}

\begin{figure}[ht!]
\centering
\includegraphics[scale=0.65]{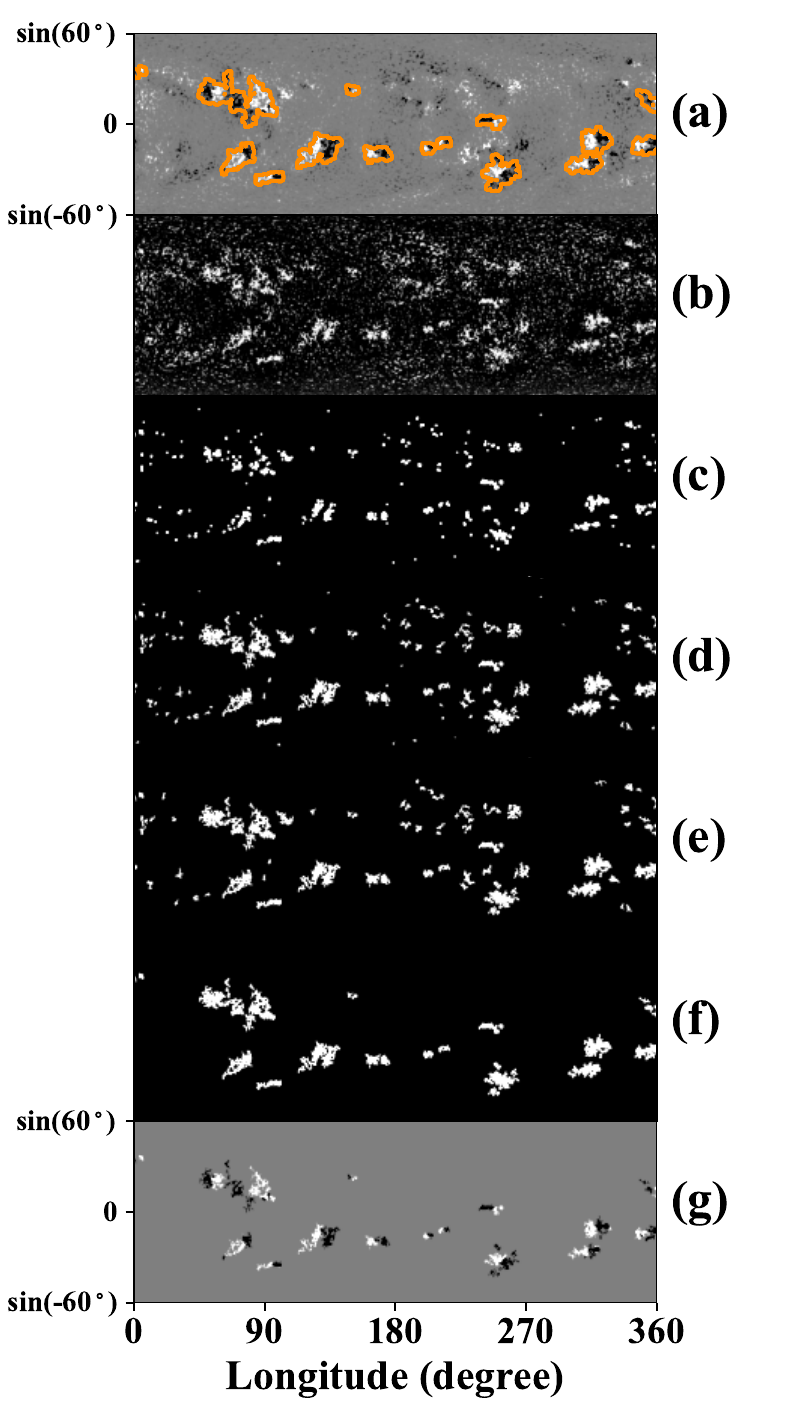}
\caption{Illustration of the AR detection algorithm. The synoptic map of CR 1968 is used as an example. Panel (a) is the original map with the ARs detection result contoured in orange lines. Panel (b): Module 1, adaptive intensity threshold segmentation; Panel (c): Module 2, morphological closing operation and opening operation; Panel (d): Module 3, region growing; Panel (e): Module 4, morphological closing operation and removing small regions; Panel (f): Module 5, merging neighbor regions and removing the unipolar regions. Panel (g) shows the final detection result of AR.
\label{fig1}}
\end{figure}

Since ARs typically emerge in middle and low latitudes and some high-latitude data from MDI and HMI magnetograms are missing, we limit our detection to the $\pm60^{\circ}$ latitude of the synoptic maps. Our AR detection algorithm operates on unsigned magnetic fields and relies on standard image processing techniques such as morphological operations and region growing. The process of AR detection is illustrated in Figure \ref{fig1}. The detection algorithm consists of five modules corresponding to Figures \ref{fig1} (b)-(f) respectively. 

The first module is to eliminate the background magnetic fields with adaptive intensity threshold segmentation. The threshold is determined by the sum of the result of Gaussian smoothing of Panel (a) and a constant value \emph{C}. The Gaussian smoothing kernel (\emph{K1}) and the value of \emph{C} are set to 501 pixels and 10 Gauss, respectively, to prevent excessive noise. Thus the threshold of each pixel varies with the magnetic field of its surrounding pixels. Panel (b) shows the resulting image. This module is highly robust and can process magnetograms with magnetic fields of varying intensities and magnetograms generated by different instruments. The module primarily removes quiet-sun magnetic fields while retaining ARs and decayed ARs.

The second module preliminarily removes the decayed ARs and identifies AR kernel pixels. It involves a morphological closing operation and an opening operation, producing the image shown in Panel (c). The size of the closing operation kernel (\emph{K2}) is set at 3 pixels while the size of the opening operation kernel (\emph{K3}) is set at 11 pixels for MDI maps and 9 pixels for HMI maps. In order to get the kernel pixels of each AR, we use the opening operation to preliminarily remove the non-ARs and some branches of ARs (non-kernel pixels). Since some small regions segmented in Panel (a) may belong to the same AR, we merge them with the closing operation to prevent their removal in the opening operation. We use the detected kernel pixels of each AR with magnetic fields greater than a certain intensity threshold as seeds in the third module. 

The third module employs region growing to obtain all pixels comprising each AR. This module identifies all pixels connected to the seeds with a field strength greater than the intensity threshold and generates Panel (d). In MDI magnetograms, the intensity threshold is set at 50 G, consistent with the threshold used in various AR detections \citep{detectionZhang, McAteerDetection, Virtanen2017, Yeates2007, MunozJaramillo2016}. For HMI magnetograms, a threshold of 30 G is chosen by trials to ensure the consistency with the detected AR area in MDI magnetograms. Module 1 uses adaptive threshold segmentation to identify nearly all ARs and segments, so region growing recovers all possible ARs. However, the segmentation also identifies some decayed ARs, part of which are still present after Module 2. Consequently, region growing also recovers these segments. As decayed ARs typically decay into small unipolar segments, the subsequent two modules remove them based on critical area and flux imbalance, respectively.

The fourth module serves to eliminate small decayed AR segments, using an area threshold (\emph{Ta}). This involves a morphological closing operation and a small region removal operation, yielding Panel (e). The area threshold (\emph{Ta}) is set at 351 pixels, equivalent to about 412 Mm$^{2}$. The magnetic flux of the smallest retained ARs is $2.42\times10^{20}$ Mx. \cite{Whitbread2018} find ARs greater than $5\times10^{20}$ Mx are enough to replicate the polar field generated by all ARs. That means the area threshold (\emph{Ta}) is sufficient for retaining small ARs whose effect on the end-of-cycle polar field can not be ignored. To avoid the removal of small regions of the same AR, a closing operation is used to merge them before the small region removal. The size of the closing operation kernel (\emph{K4}) is 5 pixels. This module removes some ephemeral regions and small decayed ARs, but the large ones are retained and will be removed in Module 5.

The fifth module is applied to remove over-decayed ARs that usually consist of unipolar regions based on their flux imbalance (\emph{Fi}) after merging neighbor regions. These unipolar regions are typically parts of decayed ARs that have been detected when they first appear on the magnetograms. They should not be detected as newly emerging flux or will affect the end-of-cycle polar field. We calculate \emph{Fi} of each AR and remove the ARs with severe flux imbalance, i.e. \emph{Fi} bigger than a threshold (\emph{Tf}). The flux imbalance \emph{Fi} is defined as the ratio of net flux to absolute flux and calculated by $Fi=(|F_+ + F_-|)/(|F_+| + |F_-|)$. The threshold \emph{Tf} is set at 0.5, consistent with the threshold in \cite{Virtanen2017,2020Yeates} for unipolar AR removal. However, some newly emerging ARs have been separated into several regions by surface flows when they appear in synoptic magnetograms. These regions may be unipolar and should not be removed simply. To avoid this, a morphological dilation operation is applied to merge neighbor regions before removing unipolar regions. The size of the dilation operation kernel (\emph{K5}) is 23 pixels. The resulting image is multiplied with Panel (e) to remove redundant areas, resulting in Panel (f). Removing unipolar regions effectively removes over-decayed AR segments, and merging neighboring regions ensures that some unipolar fragments of ARs in the early decay phase form a whole AR rather than being removed. Although some fragments are kept by connected to neighbor ARs, these two steps of module 5 effectively remove the over-decayed ARs.

Panel (f) exhibits a binary image of the identified ARs. By multiplying this panel with the initial magnetogram, Panel (a), the magnetic field distribution of the detected ARs is obtained, as demonstrated in Panel (g).

\begin{figure*}[ht!]
\centering
\includegraphics[scale=0.15]{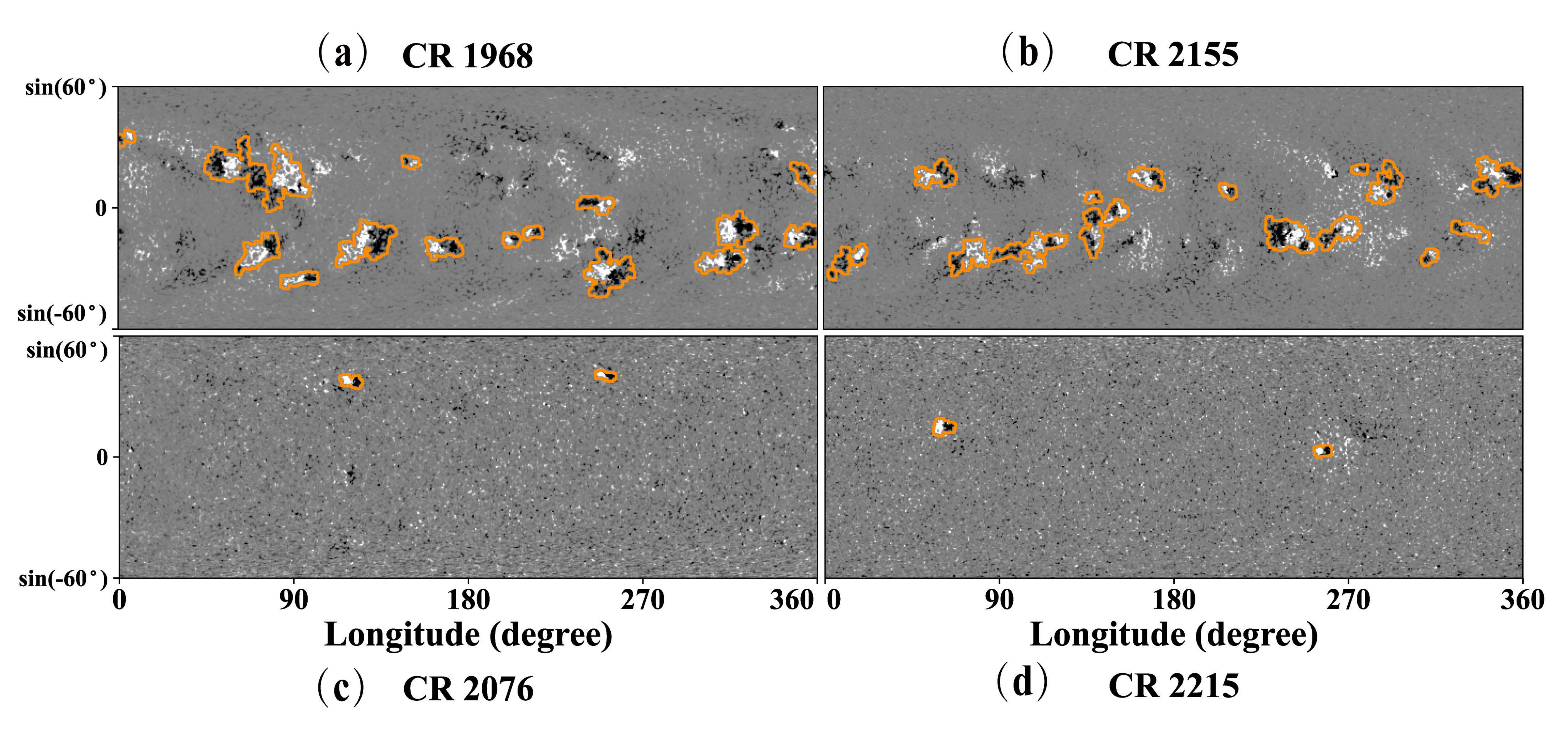}
\caption{Examples of the detected ARs based on synoptic magnetograms. 
MDI (left) and HMI (right) synoptic maps at the maximum phase (upper) and the minimum phase (lower) of cycles 23 and 24 are used. The four magnetograms are overplotted with the lines in orange outlining the profiles of the detected ARs.
\label{fig2}}
\end{figure*}

Figure \ref{fig2} shows examples of AR detections for synoptic magnetograms from MDI (left panel) and HMI (right panel) taken at the maximum phase and the minimum phase of solar cycles 23 and 24. The algorithm is able to identify all ARs in both MDI and HMI synoptic magnetograms and effectively removes most over-decayed AR segments. While the algorithm is developed using MDI and HMI synoptic maps, it can be adapted to detecting other available synoptic maps by adjusting the parameters of the detection modules.

\subsection{Controlling Parameters of the algorithm}\label{subsec:param}

\begin{table*}[!ht]
\centering
\caption{Effects of the controlling parameters in the five modules of the AR detection algorithm on the detected AR number, area, and flux. MDI synoptic magnetogram of CR 1968 is taken as an example.}
\label{table1}
\begin{threeparttable}
\begin{tabular}{cccccc}
\hline \hline
Module                   & Parameters          & Value \tnote{a}  & Number & Area (mHem) & USFlux ($10^{23}Mx$) \tnote{b}  \\ \hline
\multirow{8}{*}{Module 1} & \multirow{4}{*}{Gaussian smoothing kernel (\emph{K1})} & 101    & 14     & 50.97   & 3.523                         \\
                         &                     & 501   & 15     & 53.16   & 3.628                         \\
                         &                     & 1001   & 15     & 55.51   & 3.712                         \\
                         &                     & \emph{Diff} & 6.7\%  & 8.5\%   & 5.2\%                         \\ \cline{2-6} 
                         & \multirow{4}{*}{constant (\emph{C})}  & 5      & 16     & 55.57   & 3.727                         \\
                         &                     & 10     & 15     & 53.16   & 3.628                         \\
                         &                     & 15     & 15     & 53.11   & 3.627                         \\
                         &                     & \emph{Diff} & 7\%    & 5\%     & 3\%                           \\ \hline
\multirow{7}{*}{Module 2} & \multirow{4}{*}{closing operation kernel (\emph{K2})} & 3      & 15     & 53.16   & 3.628                         \\
                         &                     & 3      & 15     & 53.16   & 3.628                         \\
                         &                     & 5      & 17     & 57.20   & 3.792                         \\
                         &                     & \emph{Diff} & 13.3\% & 7.6\%   & 4.5\%                         \\ \cline{2-6}  
                         & \multirow{4}{*}{opening operation kernel (\emph{K3})} & 9      & 19     & 59.18   & 3.892                         \\
                         &                     & 11     & 15     & 53.16   & 3.628                         \\
                         &                     & 13     & 15     & 50.69   & 3.516                         \\
                         &                     & \emph{Diff} & 26.7\% & 16.0\%  & 10.4\%                        \\ \hline 
\multirow{8}{*}{Module 4} & \multirow{4}{*}{closing operation  kernel (\emph{K4})} & 3      & 15     & 50.88   & 3.604                         \\
                         &                     & 5      & 15     & 53.16   & 3.628                         \\
                         &                     & 7      & 16     & 56.60   & 3.688                         \\
                         &                     & \emph{Diff} & 6.7\%  & 10.8\%  & 2.3\%                         \\ \cline{2-6} 
                         & \multirow{4}{*}{area threshold (\emph{Ta})} & 251    & 15     & 53.16   & 3.628                         \\
                         &                     & 351    & 15     & 53.16   & 3.628                         \\
                         &                     & 451    & 15     & 53.00   & 3.622                         \\
                         &                     & \emph{Diff} & 0.0\%  & 0.3\%   & 0.2\%                         \\ \hline 
\multirow{8}{*}{Module 5} & \multirow{4}{*}{dilation operation kernel (\emph{K5})} & 15     & 15     & 52.50   & 3.591                         \\
                         &                     & 23     & 15     & 53.16   & 3.628                         \\
                         &                     & 31     & 15     & 53.16   & 3.628                         \\
                         &                     & \emph{Diff} & 0.0\%  & 1.3\%   & 1.0\%                         \\ \cline{2-6} 
                         & \multirow{4}{*}{flux imbalance threshold (\emph{Tf})} & 0.4    & 15     & 53.16   & 3.628                         \\
                         &                     & 0.5    & 15     & 53.16   & 3.628                         \\
                         &                     & 0.8    & 19     & 55.41   & 3.739                         \\
                         &                     & \emph{Diff} & 21.1\% & 4.1\%   & 3.0\%                         \\ \hline
\end{tabular}

\begin{tablenotes}
    \footnotesize
    \item[a] There are three values for each parameter, the lower limit, the optimized value, and the upper limit, listed from top to bottom. The acceptable range for each parameter spans from the lower limit to the upper limit. The parameter \emph{Diff} refers to the difference of detection result within the acceptable range, given by $Diff=(|V_U - V_L|)/V_O$, where $V_L, V_O, V_U$ represent the lower limit, the optimized value, and the upper limit of the three parameters, i.e. number, area, and unsigned flux. The parameters \emph{K1, K2, K3, K4, K5} and \emph{Ta} are in the unit of pixel, and \emph{C} is in the unit of Gauss.
    \item[b] Unsigned flux of detected ARs.
\end{tablenotes}

\end{threeparttable}
\end{table*}

The algorithm detects the magnetic field distribution of all ARs. For a single synoptic magnetogram, the AR magnetic field distribution is characterized by three parameters: AR number, AR area, and AR total unsigned flux. These parameters are influenced by the eight controlling parameters of the five modules, which have been introduced in Subsection \ref{subsec:alg} along with their optimized values. To evaluate the impact of each controlling parameter on the AR detection, we use the MDI synoptic magnetogram of CR 1968 as an example, and the results are presented in Table \ref{table1}. For each parameter, we provide the upper and lower limits of its acceptable range in the table. The acceptable range of each parameter is not strictly constrained here and values beyond it could also work well.

Due to the significant impact of small decayed ARs, noticeable differences (\emph{Diff}) of AR number within the acceptable range can be observed for certain parameters. However, \emph{Diff} of AR area and unsigned flux, when compared to the results obtained with the optimized value, are generally below 10\% for most parameters. By analyzing the acceptable range and \emph{Diff} of each parameter, we find that different parameters have varying effects on the algorithm. The algorithm exhibits insensitivity to many parameters, particularly \emph{Ta} in Module 4. Conversely, it displays relative sensitivity to the closing kernel (\emph{K2}) and opening kernel (\emph{K3}) in Module 2, as well as the closing kernel (\emph{K4}) in Module 4, where \emph{Diff} in AR area or flux exceeds 10\% or the acceptable ranges are small. It should be noted that although only the result of the MDI synoptic magnetogram in CR 1968 is listed in Table \ref{table1}, we have conducted additional tests using the MDI map of CR 2070 and HMI maps of CR 2155 and 2215, all of which yield similar conclusions.

The reasons for the different sensitivities of the controlling parameters are as follows. For \emph{Ta} in Module 4, it is used to remove small decayed AR segments that are typically unipolar and would be removed in Module 5 unless they are connected to other ARs. Therefore, the effect of \emph{Ta} on the results is extremely slight. In Module 2, kernel pixels of ARs are obtained while decayed ARs are removed. The closing kernel (\emph{K2}) and opening kernel (\emph{K3}) affect the distinction between ARs and decayed ARs, and the result of region growing in Module 3 further. The closing kernel (\emph{K4}) in Module 4 controls the connection of different AR segments and affects the removal of small isolated regions and unipolar regions. Since \emph{K2, K3}, and \emph{K4} not only affect the results of each operation but also strongly affect the results of the following operations, the detection result is more sensitive to them. For HMI maps, the acceptable range of \emph{K3} is 7 - 11 pixels and the optimized value is 9 pixels, different from that of the MDI maps. The reason for the difference will be given in Section \ref{sec:calibration}.

\section{Calibration and Comparison of Results from SOHO/MDI and SDO/HMI Co-temporal Magnetograms} \label{sec:calibration}

\begin{figure}[htbp!]
\centering
\includegraphics[scale=0.45]{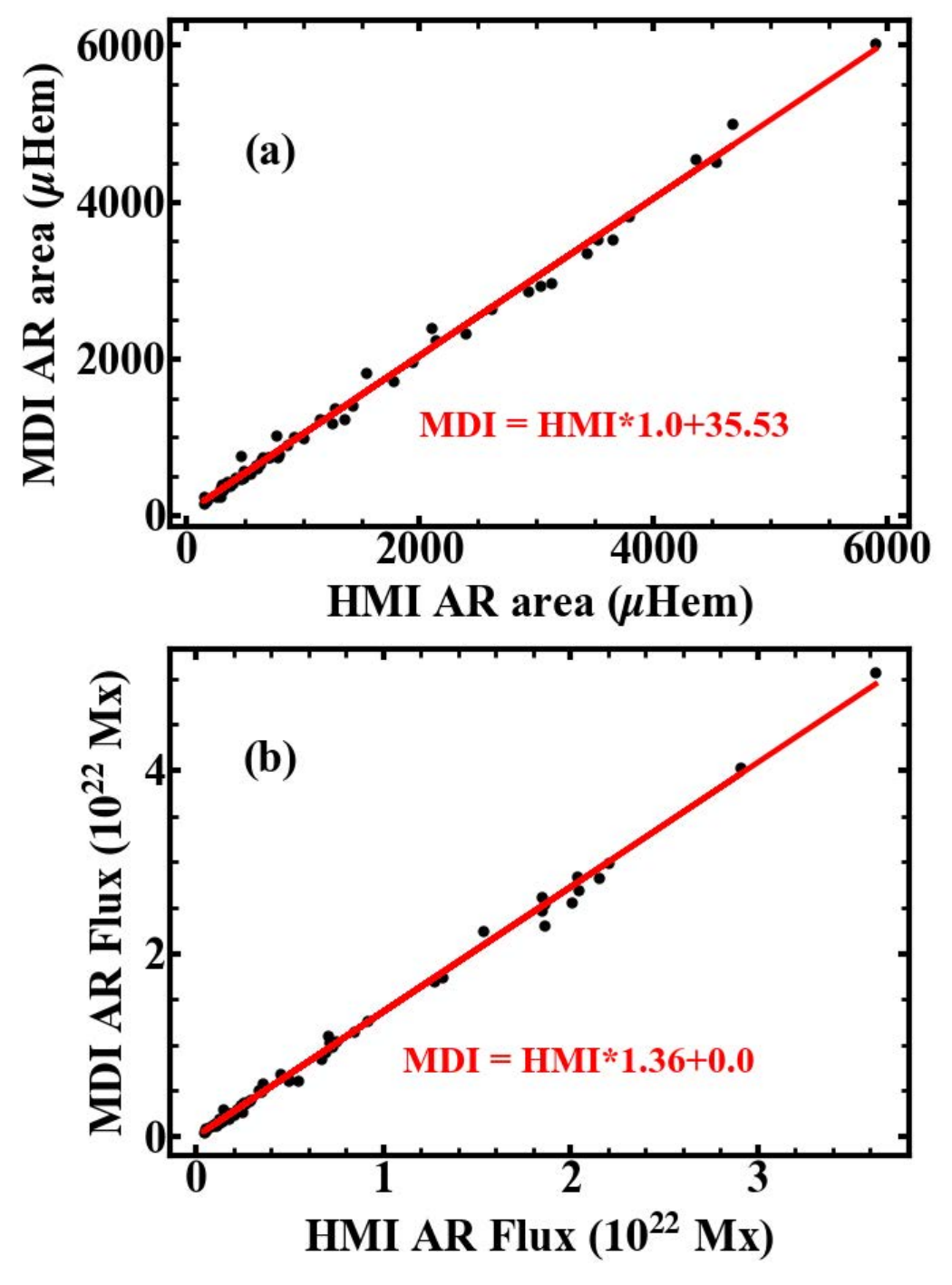}
\caption{Comparison of the area and flux of ARs detected by both MDI and HMI synoptic magnetograms during the overlap period (CRs 2097-2107). Top (Bottom): scatter plot between MDI AR area (unsigned flux) and HMI AR area (unsigned flux).
\label{fig3}}
\end{figure}

\begin{figure}[htbp!]
\centering
\includegraphics[scale=0.45]{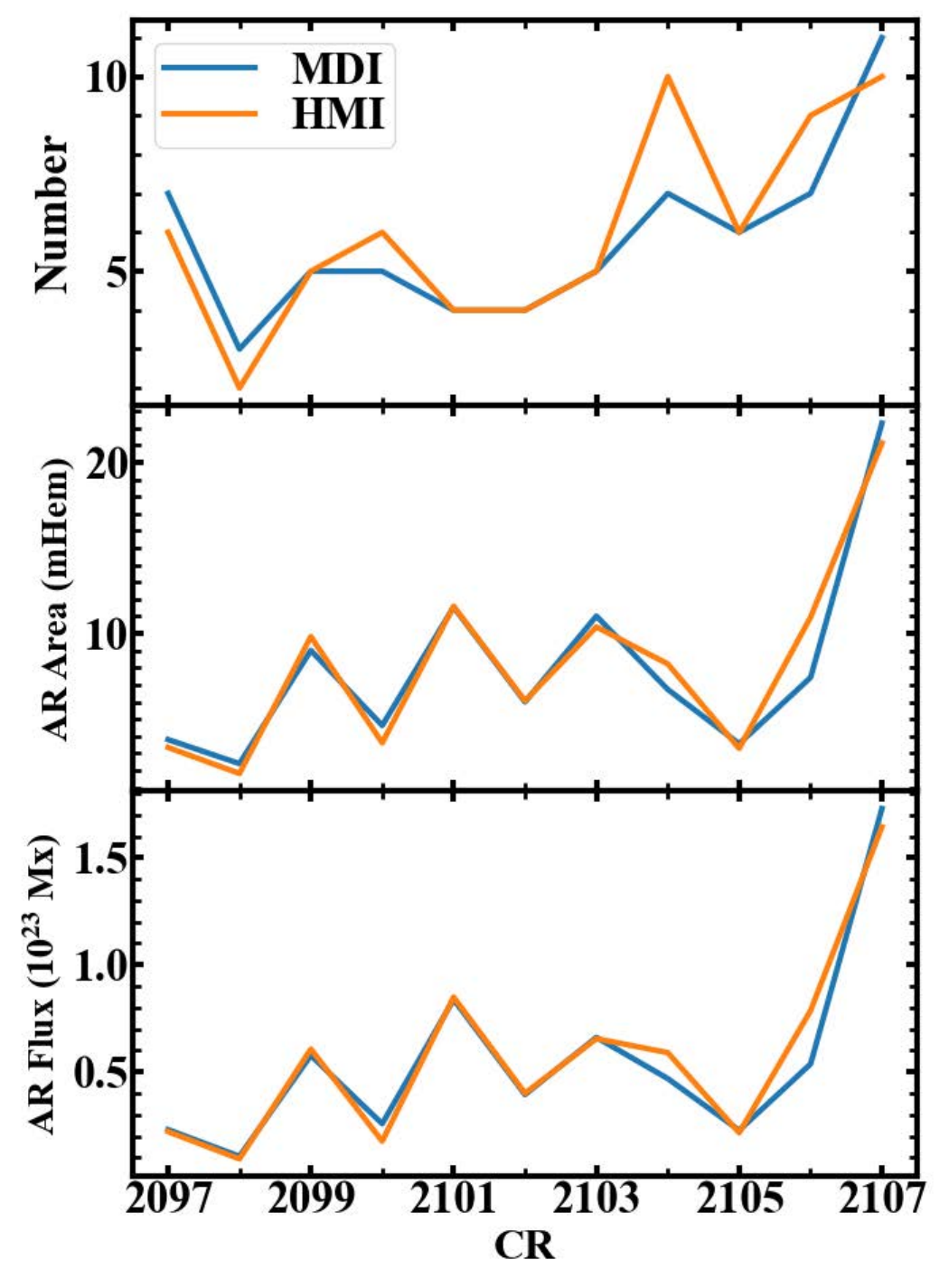}
\caption{Comparison of detected ARs during the 11 HMI and MDI overlap CRs (CRs 2097-2107) after the calibrations of detection parameters and the AR flux. From top to bottom are the evolutions of the detected AR number, area, and unsigned flux.
\label{fig4}}
\end{figure}

To ensure the homogeneity of the AR detection results obtained from MDI and HMI synoptic magnetograms, we perform two calibration processes using the co-temporal magnetograms in CRs 2097-2107. First, we calibrate the controlling parameters used for MDI and HMI maps. When using the parameters for MDI magnetograms, we find that the AR number and area detected in HMI maps are smaller than those detected in MDI maps. To obtain consistent results from the two different instruments, we reduce the opening kernel (\emph{k3}) in Module 2 from 11 pixels to 9 pixels and decrease the threshold for region growing from 50 G to 30 G. The results have been shown in Subsection \ref{subsec:alg}. After calibrating the controlling parameters, we find that the detected AR number and area of co-temporal MDI and HMI synoptic maps are consistent, which will be illustrated by Figure \ref{fig4}.

Second, we compare the area and unsigned flux of ARs detected in both MDI and HMI synoptic magnetograms to calibrate AR flux and show the effect of the parameter calibration further. In the overlap period, 67 ARs are detected in MDI maps while 64 ARs are detected in HMI maps. Different AR numbers are due to their different resolutions and magnetic field strengths, which can not be calibrated by adjusting the controlling parameters. However, we find that 56 ARs are detected in both maps, which account for over 80\% of the total ARs detected. This is similar to the 57 ARs identified by NOAA \citep{SMARPsASHARPs}.

Based on the comparison results shown in Figure \ref{fig3}, the areas detected in MDI and HMI maps are highly consistent, which shows the accuracy of the parameter calibration. However, there is a noticeable difference in the flux measurements, with MDI measurements showing a larger flux for the same AR compared to HMI measurements. The slope of the fitting function is approximately 1.36. That is similar to the results of \cite{ComparisionMDIHMI}, who find the scaling factor for fields stronger than 600 G is 1.31, for fields weaker than 600 G is 1.44, and for all pixels is 1.40.

According to the calibration above, we scale the AR flux detected in HMI maps by multiplying it with a factor of 1.36. Figure \ref{fig4} displays the results for AR number, area, and calibrated flux from co-temporal MDI and HMI magnetograms. Although the AR number differs in some CRs, the detection results for AR area and flux are highly consistent between MDI and HMI maps, except CRs 2104 and 2106. Some data are missing in MDI synoptic maps of CRs 2105 and 2106. The missing data in CR 2105 is mainly weak fields without any ARs while it contains an AR in CR 2106. As a result, no remarkable difference is found in CR 2105 but all three parameters are different in CR 2106. For CR 2104, two small ARs and one decayed AR are detected in the HMI map but not in MDI, although the effects on area and flux are slight. Figure \ref{fig4} indicates that some differences still exist for the identified result based on the MDI and HMI maps after calibration, which affects the AR number but only slightly affects the AR area and flux. We note that it is the AR area and flux that are important for our objective, that is the research of solar surface magnetic field, while the AR number is relatively insignificant.

The above calibration methods, i.e. calibration of controlling parameters and scaling of AR flux, are also required for synoptic magnetograms from other instruments when the algorithm is applied.

\section{Comparison with Other Databases}\label{sec:comparison}

Applying the automatic detection method to MDI and HMI synoptic magnetograms and calibrating the detection results between them, we generate a homogeneous database of ARs covering cycles 23, 24, and part of 25. To demonstrate the validity, homogeneity, and advantage of the method and database, we first compare the detection result of one CR map with \cite{detectionZhang} and NOAA in detail and then compare the AR number, area, and flux in cycles 23 and 24 with more databases.

\subsection{A detailed comparison of the identified results in CR 2000}\label{subsec: detailCp}

\begin{figure}[htbp!]
\centering
\includegraphics[scale=0.113]{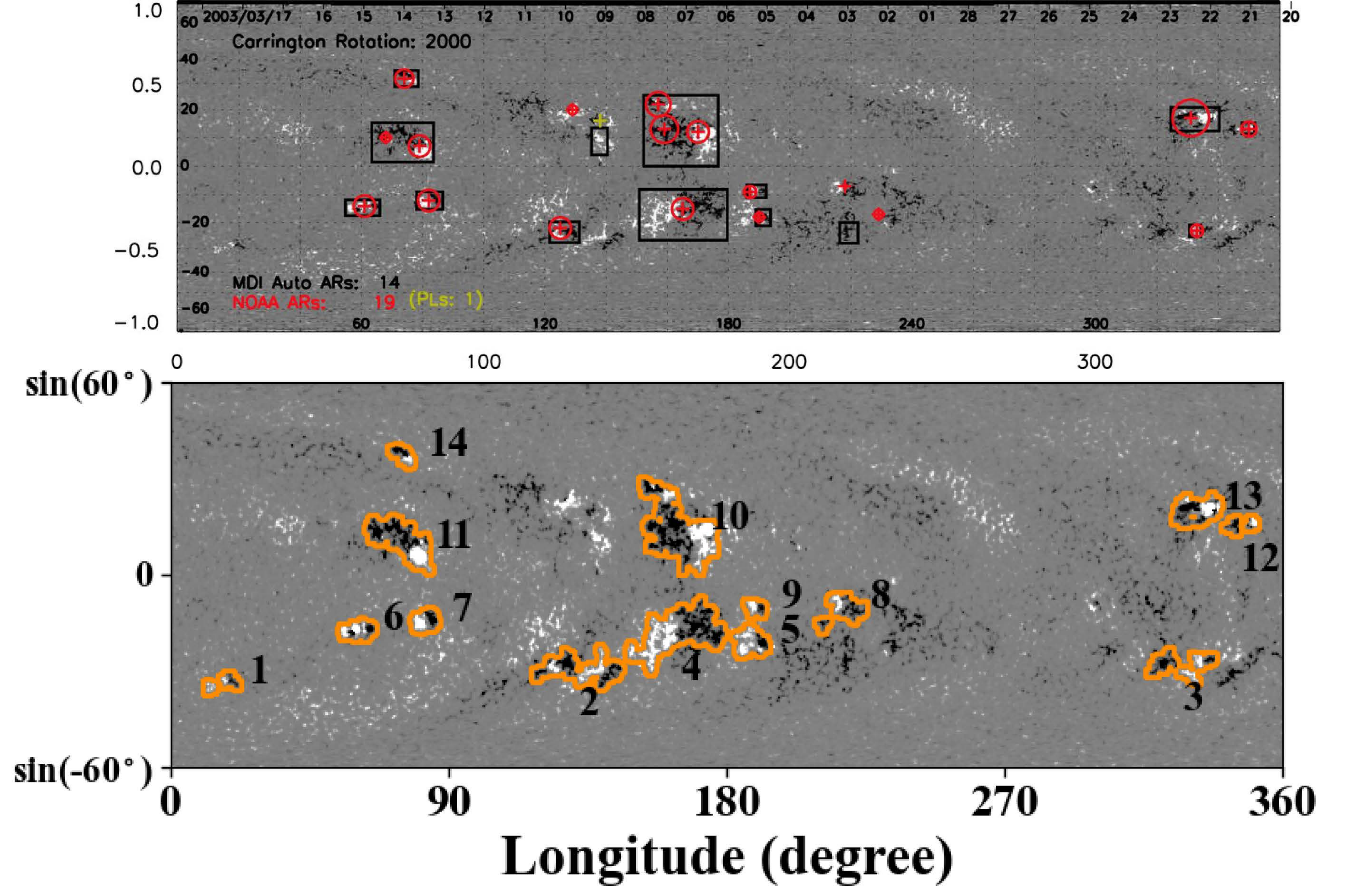}
\caption{Comparison of our results (bottom panel) with \cite{detectionZhang} and NOAA AR (top panel) of CR 2000. The top panel is reproduced with permission from \cite{detectionZhang}, copyrighted by the American Astronomical Society. ARs identified by \cite{detectionZhang} are in black boxes. NOAA ARs are labeled by red circles with plus symbols at the center indicating the centroids. The yellow symbol denotes the non-spot plage region. The bottom panel is overplotted with the contours in orange outlining the border of the detected ARs and numbers labeling them.
\label{fig5}}
\end{figure}

\cite{detectionZhang} used morphological analysis and intensity threshold to automatically identify ARs based on MDI synoptic magnetograms. They compared their AR detection result of the CR 2000 map with NOAA in the paper, which helps us to compare our detection with them and NOAA in detail.

The comparison is presented in Figure \ref{fig5}. Both we and \cite{detectionZhang} detect 14 ARs, while only 12 of them are detected by both methods. The other two ARs in \cite{detectionZhang} are unipolar regions, which are excluded by us because they are decaying ARs. For the research of solar surface magnetic field evolution, we should not include decayed ARs, which are supposed to be included in the data already when they newly emerge. Meanwhile, we detect ARs No. 1 and 8 that are not identified by them. Part of AR No. 8 is also detected by NOAA, indicating the validity of our detection. Additionally, our results are more complete than \cite{detectionZhang} for commonly detected ARs No. 3, 5, and 12. The corresponding ARs identified by \cite{detectionZhang} are nearly unipolar regions, while we detect them entirely by obtaining two polarities of each AR in Module 1 and merging them into a single one in Module 5. 

Compared with NOAA, our method detects 14 ARs while theirs detects 19 ARs. The different AR number results from two aspects. One is that some NOAA ARs are decayed ARs that are removed by us. This further indicates the property of our method. The other is that ARs No. 10 and 11 correspond to more than one NOAA AR, respectively. This means different methods to group ARs when they are crowded over the solar surface. The different methods affect the identified AR number, but not the total area and flux, which are essential parameters for our objective. AR No. 1 is not detected by NOAA. Because it has a small flux $1.82\times10^{21}$ Mx, which might not have mature sunspots \citep{Cho2015}. For our objective, such small magnetic regions are important \citep{Whitbread2018, Hofer2023}. Except for ARs No. 10, 11, and 1, most ARs detected by us correspond to one NOAA AR.

In total, our detection is similar to \cite{detectionZhang} and NOAA. However, our detection shows its advantage in properly obtaining two polarities of ARs and removing over-decayed ARs, especially unipolar regions. 

\subsection{An overall comparison of the results in cycles 23 and 24}\label{subsec: overallCp}

\begin{figure*}[htbp!]
\centering
\includegraphics[scale=0.62]{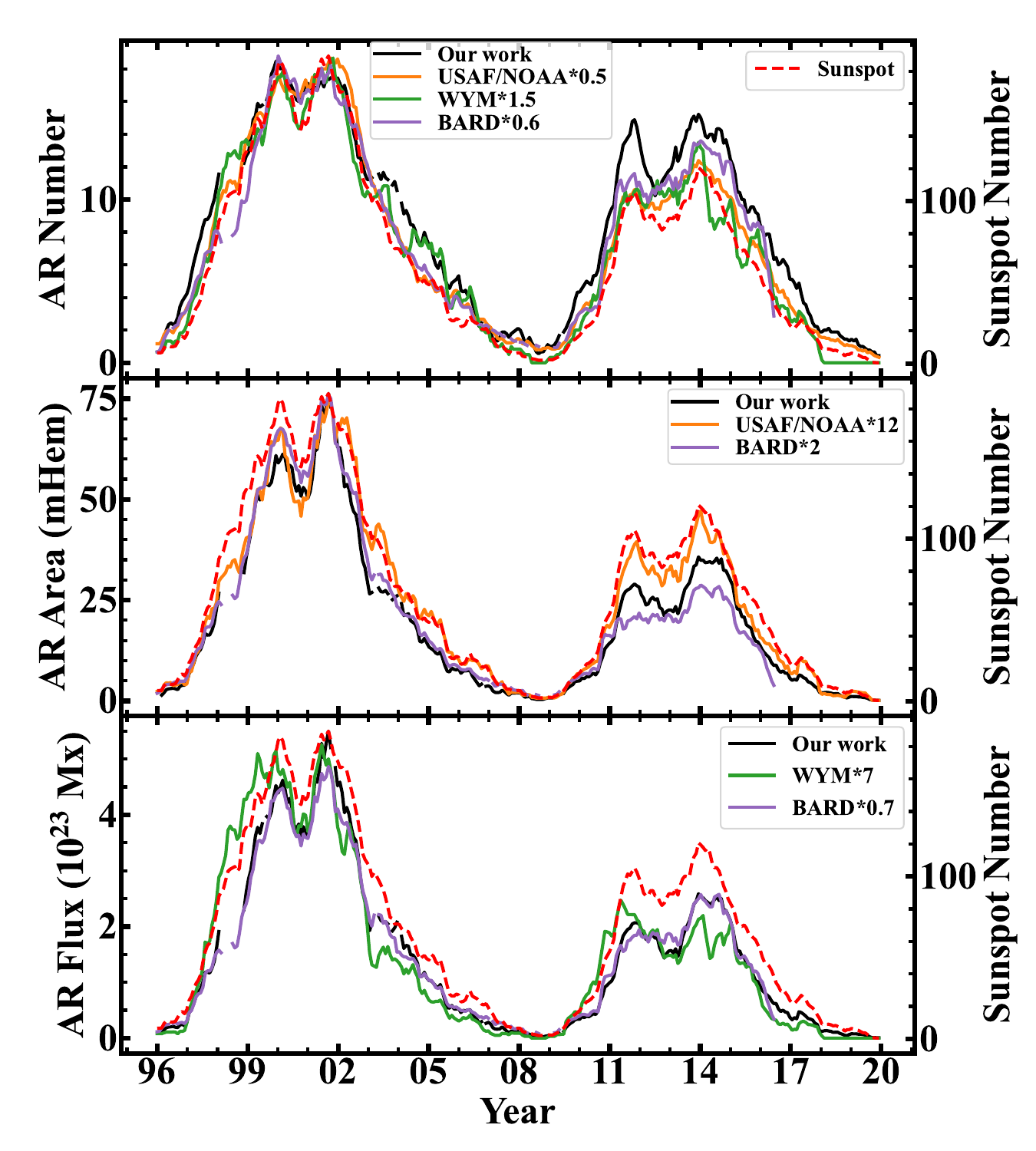}
\caption{Comparison of our results (black) with other databases in the number (top), area (middle), and flux (bottom). 13-month smoothed monthly total sunspot number: red; USAF/NOAA sunspot: orange; WYM: green; BARD: purple. Each parameter of these data is multiplied by a proper factor shown in the legend for comparison. All AR data are smoothed over nine CRs.
\label{fig6}}
\end{figure*}

To further test our database in a long time range, we compare the detection result in cycles 23 and 24 with monthly mean sunspot number, USAF/NOAA sunspot number and area, \cite{Whitbread2018} (hereafter WYM), BARD \citep{ MunozJaramillo2016, MunozJaramillo2021}, and SMARPs and SHARPs \citep{SMARPsASHARPs, Bobra2014}. USAF/NOAA observes sunspot groups daily and contains multiple records for a sunspot group, so we select each sunspot group at its maximum development of the area. Based on NSO LOS synoptic maps, WYM applies Gaussian smoothing and intensity threshold segmentation \citep{Yeates2015} to get AR properties between CR 1641 and CR 2196, covering cycles 21-24. BARD uses morphological analysis and limited human supervision on magnetograms of NSO (1996-1999), SOHO/MDI (1996-2011), and SDO/HMI (2010-2016) to detect ARs in cycles 21-24. Here we only use the BARD data of MDI and HMI. Each AR is taken when they reach the maximum flux development and the detected ARs are calibrated by multiplying AR flux in HMI with a factor of 1.25 \citep{MunozJaramillo2021}. SHARPs and SMARPs are data products from SDO/HMI and SOHO/MDI, respectively \citep{Bobra2014, SMARPsASHARPs}. The two databases provide active regions observed over the last two solar cycles, from 1996 to the present.

The comparison results are presented in Figure \ref{fig6}. All databases' AR (sunspot) number, area, and flux are calculated for each CR and smoothed with nine CRs. The gap in our data and BARD in 1998 is due to the SOHO spacecraft malfunction. The values of the three parameters of all databases in cycle 23 are adjusted to a similar strength for comparison. 

For the AR number, our database is smaller than those using full-disk maps (USAF/NOAA, BARD) because synoptic maps have a limited time resolution, which makes some small ARs that are present in full-disk magnetograms not available. However, the AR number of our database is greater than that of WYM, which also uses synoptic maps, because their method tends to merge more ARs into one AR than ours. Besides, our method is more sensitive to small ARs and can detect some small ARs that are not detected by their method. It is noted that these small ARs should not be disregarded because the impact of numerous small ARs on the end-of-cycle polar field is also significant \citep{Whitbread2018, Hofer2023}.

For the AR area, our database is greater than BARD because BARD applies a bigger threshold than us in detection, which removes AR pixels with a relatively weak magnetic field. AR area is not provided in WYM. USAF/NOAA sunspot area is severely smaller than our AR area, which is reasonable because ARs usually cover sunspots and their nearby faculae which are much larger than spots \citep{Chapman1997}.  

For the AR flux, the consistency of our database with BARD through the whole time period is highly significant, although our database is smaller in strength. One of the reasons is the different phases of AR that we choose. ARs in BARD are taken at the moment of maximum flux, while ours are taken at the moment of crossing the Central Meridian. In comparison, WYM is notably weaker than both ours and BARD in total, although it exhibits relatively stronger fluxes around 1999. This could be due to the use of different synoptic magnetograms: NSO by WYM versus MDI and HMI by us and BARD.

\begin{table}[]
\centering
\caption{Ratios of cycle 24 to cycle 23 in different parameters
}
\label{table2}
\begin{threeparttable}
\begin{tabular}{cccc}
\hline \hline
Data              & Number  & Area   & Flux   \\ \hline
\multicolumn{4}{c}{Different Databases \tnote{a}}         \\ \hline
Sunspot Number    & 63.7\%  &        &        \\
NOAA              & 66.4\%  & 63.2\% &        \\
SMARPs and SHARPs & 75.6\%  & 67.4\% & 57.3\% \\
BARD              & 72.3\%  & 37.8\% & 53.0\% \\
WYM	              & 72.3\%	&	     & 47.0\% \\
Our database          & 82.0\%  & 48.3\% & 48.4\% \\ \hline
\multicolumn{4}{c}{ARs with Different Strength \tnote{b}} \\ \hline
Strong            & 71.7\%  & 44.0\% & 45.1\% \\
Medium            & 88.1\%  & 85.3\% & 89.8\% \\
Weak              & 105.3\% & 90.9\% & 92.5\% \\ \hline
\end{tabular}

\begin{tablenotes}
    \footnotesize
    \item[a] Databases used for comparison in Subsection \ref{subsec: overallCp}
    \item[b] ARs with different strengths of our database. They are described in detail in Subsection \ref{subsec: ARstrength}.
\end{tablenotes}

\end{threeparttable}
\end{table}

To assess the homogeneity of our database, we conduct a comparative analysis of the ratios of cycle 24 to cycle 23 for the three parameters among the aforementioned databases. The ratios are calculated using the peak values in two cycles for all parameters and they are presented in Table \ref{table2}. Although Figure \ref{fig6} does not display the ratios for SMARPs and SHARPs, we include their ratios in Table \ref{table2} to ensure a comprehensive comparison.

We find the ratio of AR number in our database surpasses that of the other databases. This discrepancy can be attributed to the merging of neighboring regions in Module 5, which serves to combine separate regions into a single active region but also merges closely located active regions into one. Given the considerably stronger activity during cycle 23, a larger number of active regions merge, resulting in a slightly higher ratio. On the other hand, the ratios of AR area and flux in our database are found to be situated in the middle range among the ratios given by other databases and the ratios of area and flux are almost identical. This similarity is consistent with the well-known linear correlation between AR area and AR flux \citep{Sheeley1966}, implying the homogeneity of our database in terms of area and flux. Compared to the sunspot number, our database exhibits a higher ratio of AR number, while the ratios of AR area and flux are comparatively lower. The other databases also exhibit similar results, except NOAA for their detection according to sunspots. Furthermore, the ratio of AR number exceeds the ratios of AR area and flux both in our database and other databases.

\subsection{Analysis of different ratios of cycle 24 to cycle 23 in different parameters}\label{subsec: ARstrength}

\begin{figure}[htbp!]
\centering
\includegraphics[scale=0.43]{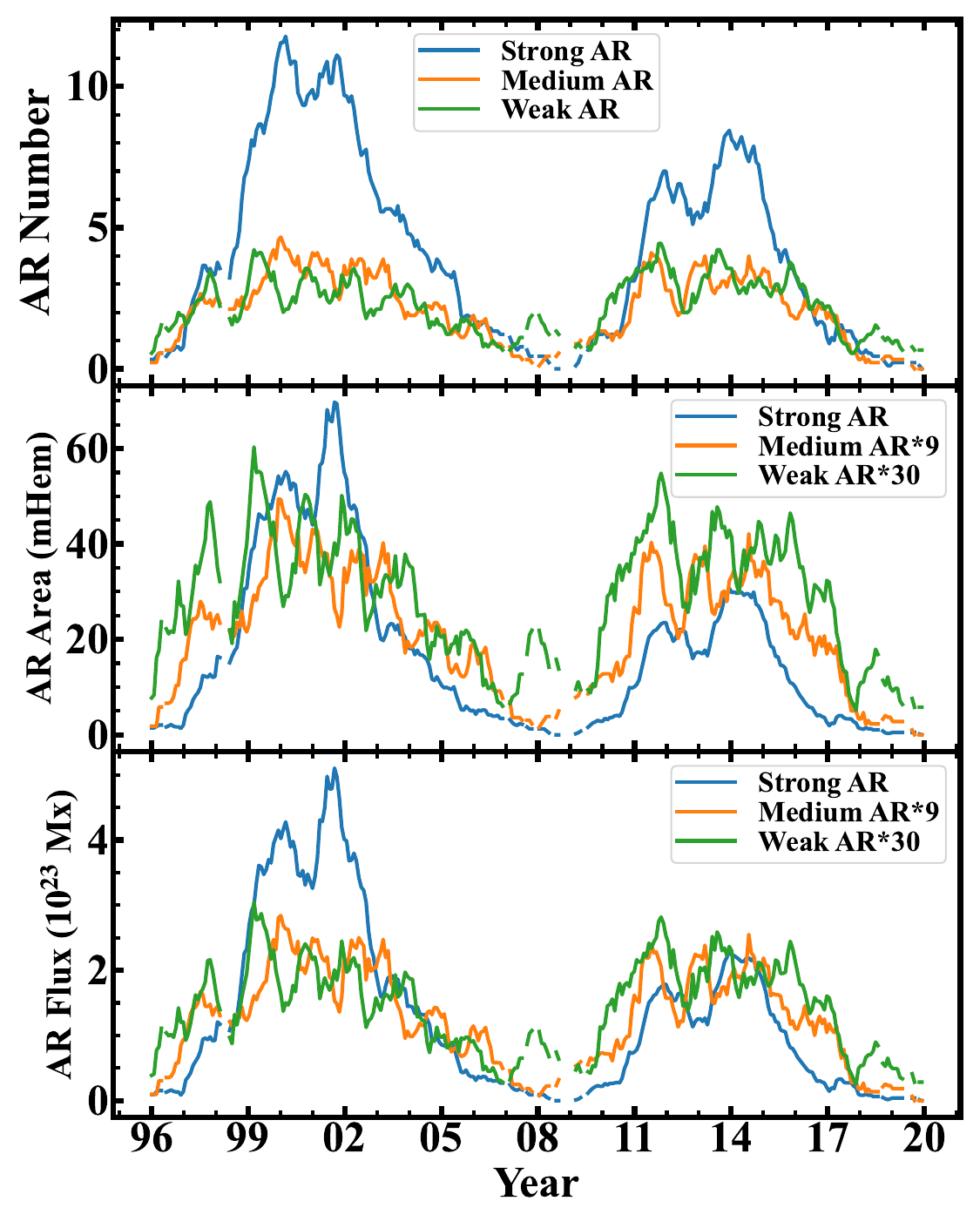}
\caption{Statistical properties (Number: top; Area: middle; Unsigned Flux: bottom) of the detected ARs with different strengths of flux in cycles 23 and 24. The area and flux of medium AR and weak AR are multiplied with a factor shown in the legend for comparison with strong AR.
\label{fig7}}
\end{figure}

To understand the cause of different ratios in number and area and whether they suggest the inhomogeneity of our database, we analyze ARs with different strengths in cycles 23 and 24. According to the unsigned flux of ARs, we separate ARs into three categories, strong ARs ($|flux| > 10^{22}Mx$), medium ARs ($4\times10^{21} < |flux| < 10^{22}Mx$), and weak ARs ($|flux| < 4\times10^{21}$). The thresholds for ARs of different strengths are set according to \cite{WangYM1989}. Among 2579 ARs, there are 1266 strong ARs, 640 medium ARs, and 673 weak ARs. 

The number, area, and flux of the three categories are shown in Figure 7 and the ratios of cycle 24 to cycle 23 in different parameters are shown in Table 2. The area ratios keep the same as the ratios of the flux for all three categories, but the difference between the area and number varies for them. The ratios of area and number exhibit significant disparities for strong ARs, while they are similar for medium and weak ARs. In terms of area and flux, cycle 24 is even more than half weaker than that of cycle 23 for the strong ARs. With the decrease of the ARs’ strength, the difference between the two cycles in area and flux decreases. For weak ARs’ area and flux, cycle 24 is almost the same as that of cycle 23. The number ratios demonstrate a similar trend. Cycle 24 even has slightly more weak ARs than cycle 23. In total, the ratio of total AR numbers is 82\% which is between the ratios of strong ARs and weak ARs. However, the ratios of total AR area and unsigned flux are similar to that of strong ARs because weak and medium ARs contribute little to the total area and flux. 

\cite{Toma2013} also find there is a notable decrease in the number of large sunspots during cycle 23 compared to cycle 22. In contrast, the numbers of small sunspots remain relatively consistent between the two cycles. Our results, in alignment with the findings of \cite{Toma2013}, indicate that the variations across different solar cycles are primarily driven by the strong active regions (large sunspots), while weak active regions (small sunspots) exhibit minimal changes in response to cycle strength. Moreover, our results indicate that the different ratios in terms of AR area and number between cycle 24 and cycle 23 are primarily affected by the strong ARs and are unlikely to signify inhomogeneity within our database. They are the intrinsic property of the solar cycle, which implies that the relative strength of solar cycles varies with different activity indices. The widely adopted cycle strength is based on the synthetic index of sunspot number. We will further verify the index dependence of cycle strength using the sunspot area data since 1874 and other historical datasets in a forthcoming study.

\section{Basic AR parameters of the database} \label{sec:parameters}

The automatic detection method detects the magnetic field distribution of each AR. Based on that, parameters characterizing each AR can be calculated. As the first paper, we just provide several basic parameters, including the latitude and longitude of the flux-weighted centroid of two polarities and the whole AR, area, and flux of each polarity. The CR number and label jointly serve to identify a unique AR. Table \ref{table3} lists these parameters. More parameters, including equivalent polarity separation and tilt angle when each AR is approximated as a BMR \citep{2020Yeates}, initial dipole moment, final dipole moment, and ARDoR, will be given in the second paper of the series.

\begin{figure}[htbp!]
\centering
\includegraphics[scale=0.34]{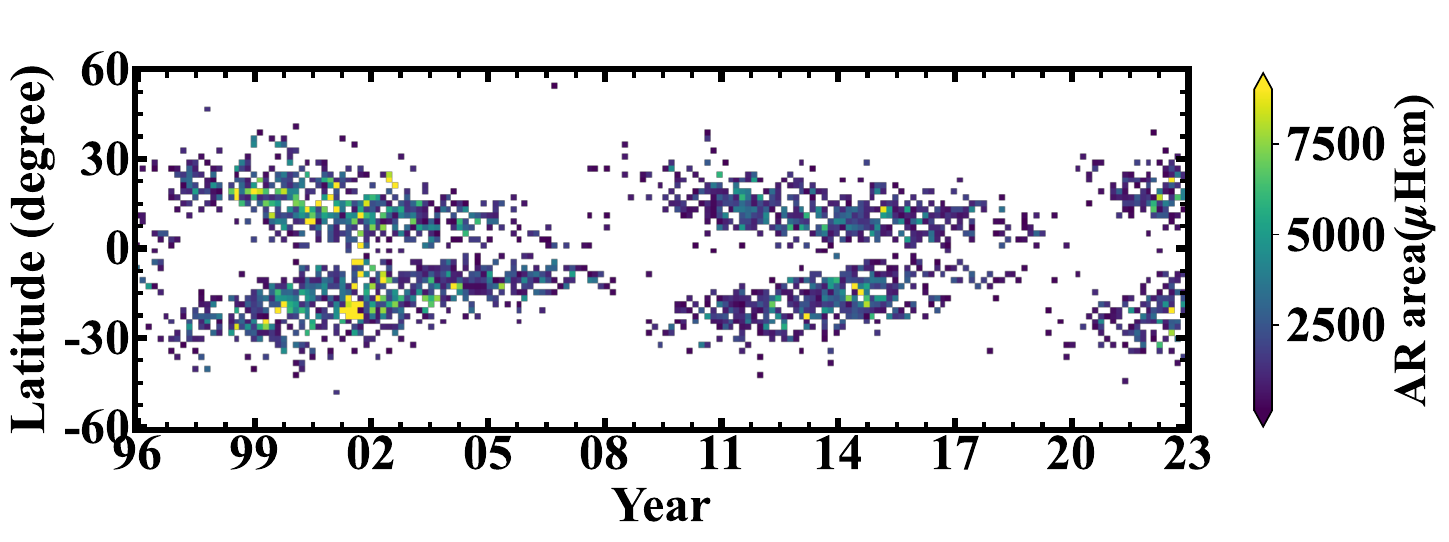}
\caption{Butterfly diagram of ARs of our database in cycles 23, 24, and part of 25. The color shows the average area of ARs. There is a gap in 1998 due to the missing synoptic magnetograms in CRs 1938-1940. 
\label{fig8}}
\end{figure}

Our database currently covers the time range from May 5, 1996, to January 1, 2023 (CRs 1909 - 2265) and includes cycles 23, 24, and part of 25, with continuous extensions planned. Figure \ref{fig8} displays the latitude of ARs versus time from 1996 to 2023. We detect 2849 ARs, with 1481 ARs in cycle 23 and 1098 ARs in cycle 24. The detected ARs in cycle 23 are less than the 1730 ARs detected by \cite{detectionZhang} because they keep some unipolar ARs in their detection. The number of ARs in cycle 24 is similar to the results of \cite{2020Yeates}, who detects 1090 BMRs using SHARPs data from May 2010 to April 2020, approximately during cycle 24 (December 2008 to January 2020). The average area and unsigned flux of ARs are 2421 $\mu$Hem and $1.75\times10^{22}$ Mx, respectively. The smallest AR in terms of area is 136 $\mu$Hem with a flux of $4.16\times10^{20}$ Mx, weaker than $5\times10^{20}$ Mx, able to retain enough small ARs to reconstruct the end-of-cycle polar field \citep{Whitbread2018}. The largest AR detected in our database is 30122 $\mu$Hem with a flux of $1.84\times10^{23}$ Mx.

The detection algorithm and full database of detected ARs are freely accessible in \href{https://github.com/Wang-Ruihui/A-live-homogeneous-database-of-solar-active-regions}{https://github.com/Wang-Ruihui/A-live-homogeneous-database-of-solar-active-regions}.

\begin{longrotatetable}
\begin{table*}[]
\begin{threeparttable}
\centering
\caption{AR database \tnote{a}.
}
\label{table3}
\setlength{\tabcolsep}{1mm}{
\begin{tabular}{cccccccccccc}
\hline \hline
CR   & Label & Lat (+) \tnote{b} & Lon (+) & Lat (-) \tnote{c} & Lon (-) & Lat (whole) \tnote{d} & Lon (whole) & Area (+, $\mu$Hem) & Area (-, $\mu$Hem) & Flux (+, Mx) & Flux (-, Mx) \\ \hline
1968 & 1     & -26.95   & 249.36   & -28.62   & 254.66   & -27.93     & 252.46     & 2522.38         & 3414.35         & 1.51E+22      & -2.13E+22     \\
1968 & 2     & -31.02   & 91.57    & -30.26   & 99.38    & -30.61     & 95.76      & 1081.40         & 658.56          & 5.99E+21      & -6.95E+21     \\
1968 & 3     & -19.91   & 71.51    & -16.75   & 76.92    & -18.51     & 73.91      & 2396.22         & 1638.89         & 1.23E+22      & -9.79E+21     \\
1968 & 4     & -22.61   & 308.29   & -21.71   & 316.41   & -22.19     & 312.06     & 2234.18         & 1606.48         & 1.70E+22      & -1.47E+22     \\
1968 & 5     & -14.47   & 123.95   & -13.62   & 134.23   & -14.04     & 129.20     & 3474.92         & 3306.33         & 2.12E+22      & -2.21E+22     \\
1968 & 6     & -16.67   & 163.48   & -16.34   & 171.97   & -16.49     & 168.24     & 1211.42         & 1303.63         & 7.59E+21      & -9.68E+21     \\
1968 & 7     & -11.89   & 347.88   & -11.51   & 356.03   & -11.70     & 351.87     & 1530.86         & 1683.64         & 1.42E+22      & -1.36E+22     \\
1968 & 8     & -13.57   & 200.86   & -13.00   & 204.31   & -13.33     & 202.31     & 407.79          & 420.52          & 4.00E+21      & -2.89E+21     \\
1968 & 9     & -9.56    & 314.53   & -8.22    & 322.46   & -8.86      & 318.67     & 2744.60         & 2273.53         & 2.00E+22      & -2.17E+22     \\
1968 & 10    & -10.27   & 211.27   & -9.74    & 215.42   & -10.03     & 213.17     & 332.95          & 342.98          & 2.70E+21      & -2.28E+21     \\
1968 & 11    & 1.40     & 251.13   & 2.17     & 242.65   & 1.75       & 247.33     & 826.00          & 1063.27         & 8.26E+21      & -6.71E+21     \\
1968 & 12    & 14.29    & 77.02    & 14.32    & 64.93    & 14.30      & 71.30      & 7381.56         & 6999.61         & 4.70E+22      & -4.22E+22     \\
1968 & 13    & 11.77    & 358.77   & 13.54    & 352.97   & 12.72      & 355.65     & 462.19          & 856.48          & 3.67E+21      & -4.28E+21     \\
1968 & 14    & 18.54    & 151.98   & 19.33    & 148.39   & 18.87      & 150.50     & 297.84          & 168.98          & 1.37E+21      & -9.64E+20     \\
1968 & 15    & 30.88    & 5.86     & 29.26    & 0.91     & 30.33      & 4.18       & 330.63          & 192.13          & 2.04E+21      & -1.05E+21     \\ \hline
\end{tabular}}

\begin{tablenotes}
    \footnotesize
    \item[a] Table 3 is published in its entirety in the machine-readable format. The ARs in the CR 1968 synoptic map are shown here for guidance regarding its form and content.
    \item[b] All longitudes and latitudes are in the unit of degree. The symbol `+' refers to the positive polarity of the AR.
    \item[c] The symbol `-' refers to the negative polarity of the AR.
    \item[d] `Wholes' refers to the whole AR.
\end{tablenotes}

\end{threeparttable}
\end{table*}
\end{longrotatetable}

\section{Conclusion and Discussion} \label{sec:conclusion}
In order to provide a homogeneous AR database for the understanding and prediction of the solar cycle, we propose a new method to automatically detect ARs from MDI and HMI synoptic magnetograms, calibrate the detections from MDI and HMI maps, and provide several basic parameters of the detected ARs.

Our method for AR detection is based on morphological operations and region growing. It can process synoptic magnetograms with varying magnetic field strengths and maps from different instruments. It is able to identify all possible ARs and remove unipolar regions. Unipolar regions are typically part of decayed ARs and should be excluded from our database. Otherwise, they will erroneously impact the analysis of the end-of-cycle polar field. Compared to \cite{detectionZhang} and NOAA, our detection is similar to them overall but our method shows its advantage in properly detecting two polarities of ARs and removing the unipolar regions.

To obtain a homogeneous AR database, we apply calibrations to the detections from MDI and HMI synoptic magnetograms, specifically adjusting the controlling parameters and the detected unsigned flux. Through a comparative analysis of ARs detected on both maps during the overlap period, we find that the AR flux in MDI maps is approximately 1.36 times that of HMI maps, consistent with the calibration of \cite{ComparisionMDIHMI}. 

When compared to other databases such as the sunspot number, USAF/NOAA sunspot number and area, \cite{Whitbread2018}, BARD, SMARPs, and SHARPs, our database exhibits a similar trend of the time evolution of AR numbers, areas, and unsigned flux in cycles 23 and 24. However, the ratios of cycle 24 to cycle 23 differ among these databases for all three parameters. Specifically, our database and most others exhibit a higher ratio of AR number compared to the widely used sunspot number, while the ratios of AR area and flux are relatively lower. Additionally, the ratio of AR number consistently surpasses the ratios of AR area and flux across our database and the other databases. Through our analysis of ARs with different strengths in cycles 23 and 24, we find that the distinct ratios in AR number, area, and flux are primarily influenced by the strong ARs, while the weak ARs show similar ratios for the AR number and area. This indicates that the different ratios are not caused by the inhomogeneity within our database, but show that the strength of the solar cycle varies with different indices of solar activity. Furthermore, our analysis suggests that weaker ARs exhibit weaker dependence on the solar cycle, and the difference in the strength of cycles 23 and 24 is primarily caused by strong active regions. 

Although several rogue ARs significantly affect the end-of-cycle polar field, the effect of small ARs can not be ignored. Their contribution to the polar field is even comparable to that of big ARs \citep{Hofer2023}. However, according to the study of \cite{Whitbread2018}, ARs greater than $5\times10^{20}$ Mx are enough to replicate the polar field generated by all ARs. The weakest AR in our database is $2.42\times10^{20}$ Mx. It means that our database contains enough small ARs for the research of surface magnetic field evolution and solar cycle prediction. Besides the advantages presented above, there are still limitations of our database so far. The ARs in our detection are not in their fully emerged phase. It is attributed to the time resolution of synoptic magnetograms and the absence of the observation of the far side of the sun. Some ARs are detected repeatedly because they appear in several synoptic maps due to their long lifetimes. In addition, our database just presents several basic parameters now. In the subsequent study, we will remove the repeated ARs properly, and provide and analyze more useful parameters, particularly the final dipole field that quantifies the impact of ARs on the end-of-cycle polar field. Additionally, we will continuously update the database based on newly released synoptic magnetograms.\\ \hspace*{\fill} \\

We thank the referee for the valuable comments and suggestions on improving the paper. The research is supported by the National Natural Science Foundation of China No. 12173005 and National Key R\&D Program of China No. 2022YFF0503800. J.J. acknowledges the International Space Science Institute Teams 474. The SDO/HMI data are courtesy of NASA and the SDO/HMI team. SOHO is a project of international cooperation between ESA and NASA. Sunspot data are from the World Data Center SILSO, Royal Observatory of Belgium, Brussels. Active region dipole moments determined from NSO synoptic carrington maps were downloaded from the solar dynamo dataverse (https://dataverse.harvard.edu/dataverse/solardynamo), maintained by Andrés Muñoz-Jaramillo. 


\bibliography{sample631}{}

\begin{thebibliography}{}
\expandafter\ifx\csname natexlab\endcsname\relax\def\natexlab#1{#1}\fi
\providecommand{\url}[1]{\href{#1}{#1}}
\providecommand{\dodoi}[1]{doi:~\href{http://doi.org/#1}{\nolinkurl{#1}}}
\providecommand{\doeprint}[1]{\href{http://ascl.net/#1}{\nolinkurl{http://ascl.net/#1}}}
\providecommand{\doarXiv}[1]{\href{https://arxiv.org/abs/#1}{\nolinkurl{https://arxiv.org/abs/#1}}}

\bibitem[{{Babcock}(1961)}]{Babcock1961}
{Babcock}, H.~W. 1961, \apj, 133, 572, \dodoi{10.1086/147060}

\bibitem[{{Bobra} {et~al.}(2014){Bobra}, {Sun}, {Hoeksema}, {Turmon}, {Liu},
  {Hayashi}, {Barnes}, \& {Leka}}]{Bobra2014}
{Bobra}, M.~G., {Sun}, X., {Hoeksema}, J.~T., {et~al.} 2014, \solphys, 289,
  3549, \dodoi{10.1007/s11207-014-0529-3}

\bibitem[{{Bobra} {et~al.}(2021){Bobra}, {Wright}, {Sun}, \&
  {Turmon}}]{SMARPsASHARPs}
{Bobra}, M.~G., {Wright}, P.~J., {Sun}, X., \& {Turmon}, M.~J. 2021, \apjs,
  256, 26, \dodoi{10.3847/1538-4365/ac1f1d}

\bibitem[{{Chapman} {et~al.}(1997){Chapman}, {Cookson}, \&
  {Dobias}}]{Chapman1997}
{Chapman}, G.~A., {Cookson}, A.~M., \& {Dobias}, J.~J. 1997, \apj, 482, 541,
  \dodoi{10.1086/304138}

\bibitem[{{Cho} {et~al.}(2015){Cho}, {Cho}, {Bong}, {Lim}, {Kim}, {Choi},
  {Kim}, \& {Yurchyshyn}}]{Cho2015}
{Cho}, I.~H., {Cho}, K.~S., {Bong}, S.~C., {et~al.} 2015, \apj, 811, 49,
  \dodoi{10.1088/0004-637X/811/1/49}

\bibitem[{{de Toma} {et~al.}(2013){de Toma}, {Chapman}, {Preminger}, \&
  {Cookson}}]{Toma2013}
{de Toma}, G., {Chapman}, G.~A., {Preminger}, D.~G., \& {Cookson}, A.~M. 2013,
  \apj, 770, 89, \dodoi{10.1088/0004-637X/770/2/89}

\bibitem[{{Hale} {et~al.}(1919){Hale}, {Ellerman}, {Nicholson}, \&
  {Joy}}]{Hale1919}
{Hale}, G.~E., {Ellerman}, F., {Nicholson}, S.~B., \& {Joy}, A.~H. 1919, \apj,
  49, 153, \dodoi{10.1086/142452}

\bibitem[{{Hofer} {et~al.}(2023){Hofer}, {Krivova}, {Cameron}, {Solanki}, \&
  {Jiang}}]{Hofer2023}
{Hofer}, B., {Krivova}, N.~A., {Cameron}, R., {Solanki}, S.~K., \& {Jiang}, J.
  2023, \aap, under view, Ax, \dodoi{xxx}

\bibitem[{{Jaeggli} \& {Norton}(2016)}]{Jaeggli2016}
{Jaeggli}, S.~A., \& {Norton}, A.~A. 2016, \apjl, 820, L11,
  \dodoi{10.3847/2041-8205/820/1/L11}

\bibitem[{{Jiang} {et~al.}(2011){Jiang}, {Cameron}, {Schmitt}, \&
  {Sch{\"u}ssler}}]{Jiang2011}
{Jiang}, J., {Cameron}, R.~H., {Schmitt}, D., \& {Sch{\"u}ssler}, M. 2011,
  \aap, 528, A82, \dodoi{10.1051/0004-6361/201016167}

\bibitem[{{Jiang} {et~al.}(2014){Jiang}, {Cameron}, \&
  {Sch{\"u}ssler}}]{Jiang2014}
{Jiang}, J., {Cameron}, R.~H., \& {Sch{\"u}ssler}, M. 2014, \apj, 791, 5,
  \dodoi{10.1088/0004-637X/791/1/5}

\bibitem[{{Jiang} {et~al.}(2015){Jiang}, {Cameron}, \&
  {Sch{\"u}ssler}}]{Jiang2015}
---. 2015, \apjl, 808, L28, \dodoi{10.1088/2041-8205/808/1/L28}

\bibitem[{{Jiang} {et~al.}(2007){Jiang}, {Chatterjee}, \&
  {Choudhuri}}]{Jiang2007}
{Jiang}, J., {Chatterjee}, P., \& {Choudhuri}, A.~R. 2007, \mnras, 381, 1527,
  \dodoi{10.1111/j.1365-2966.2007.12267.x}

\bibitem[{{Jiang} {et~al.}(2019){Jiang}, {Song}, {Wang}, \&
  {Baranyi}}]{Jiang2019}
{Jiang}, J., {Song}, Q., {Wang}, J.-X., \& {Baranyi}, T. 2019, \apj, 871, 16,
  \dodoi{10.3847/1538-4357/aaf64a}

\bibitem[{{K{\"u}nzel}(1960)}]{Kunzel1960}
{K{\"u}nzel}, H. 1960, Astronomische Nachrichten, 285, 271,
  \dodoi{10.1002/asna.19592850516}

\bibitem[{{Leighton}(1969)}]{Leighton1969}
{Leighton}, R.~B. 1969, \apj, 156, 1, \dodoi{10.1086/149943}

\bibitem[{{Liu} {et~al.}(2012){Liu}, {Hoeksema}, {Scherrer}, {Schou},
  {Couvidat}, {Bush}, {Duvall}, {Hayashi}, {Sun}, \&
  {Zhao}}]{ComparisionMDIHMI}
{Liu}, Y., {Hoeksema}, J.~T., {Scherrer}, P.~H., {et~al.} 2012, \solphys, 279,
  295, \dodoi{10.1007/s11207-012-9976-x}

\bibitem[{{McAteer} {et~al.}(2005){McAteer}, {Gallagher}, {Ireland}, \&
  {Young}}]{McAteerDetection}
{McAteer}, R.~T.~J., {Gallagher}, P.~T., {Ireland}, J., \& {Young}, C.~A. 2005,
  \solphys, 228, 55, \dodoi{10.1007/s11207-005-4075-x}

\bibitem[{{Mu{\~n}oz-Jaramillo} {et~al.}(2013){Mu{\~n}oz-Jaramillo},
  {Dasi-Espuig}, {Balmaceda}, \& {DeLuca}}]{Munoz-Jaramillo2013}
{Mu{\~n}oz-Jaramillo}, A., {Dasi-Espuig}, M., {Balmaceda}, L.~A., \& {DeLuca},
  E.~E. 2013, \apjl, 767, L25,
  \dodoi{10.1088/2041-8205/767/2/L2510.48550/arXiv.1304.3151}

\bibitem[{{Mu{\~n}oz-Jaramillo} {et~al.}(2021){Mu{\~n}oz-Jaramillo},
  {Navarrete}, \& {Campusano}}]{MunozJaramillo2021}
{Mu{\~n}oz-Jaramillo}, A., {Navarrete}, B., \& {Campusano}, L.~E. 2021, \apj,
  920, 31, \dodoi{10.3847/1538-4357/ac133b}

\bibitem[{Muñoz-Jaramillo {et~al.}(2016)Muñoz-Jaramillo, Werginz,
  Vargas-Acosta, DeLuca, Windmueller, Zhang, Longcope, Lamb, DeForest,
  Vargas-Domínguez, Harvey, \& Martens}]{MunozJaramillo2016}
Muñoz-Jaramillo, A., Werginz, Z.~A., Vargas-Acosta, J.~P., {et~al.} 2016, in
  2016 IEEE International Conference on Big Data (Big Data), 3194--3203,
  \dodoi{10.1109/BigData.2016.7840975}

\bibitem[{{Nagy} {et~al.}(2017){Nagy}, {Lemerle}, {Labonville}, {Petrovay}, \&
  {Charbonneau}}]{Nagy2017}
{Nagy}, M., {Lemerle}, A., {Labonville}, F., {Petrovay}, K., \& {Charbonneau},
  P. 2017, \solphys, 292, 167, \dodoi{10.1007/s11207-017-1194-0}

\bibitem[{{Nagy} {et~al.}(2020){Nagy}, {Petrovay}, {Lemerle}, \&
  {Charbonneau}}]{Nagy2020}
{Nagy}, M., {Petrovay}, K., {Lemerle}, A., \& {Charbonneau}, P. 2020, Journal
  of Space Weather and Space Climate, 10, 46, \dodoi{10.1051/swsc/2020051}

\bibitem[{{Petrovay}(2020)}]{Petrovay2020}
{Petrovay}, K. 2020, Living Reviews in Solar Physics, 17, 2,
  \dodoi{10.1007/s41116-020-0022-z}

\bibitem[{{Petrovay} {et~al.}(2020){Petrovay}, {Nagy}, \&
  {Yeates}}]{Petrovay2020b}
{Petrovay}, K., {Nagy}, M., \& {Yeates}, A.~R. 2020, Journal of Space Weather
  and Space Climate, 10, 50, \dodoi{10.1051/swsc/2020050}

\bibitem[{{Schatten} {et~al.}(1978){Schatten}, {Scherrer}, {Svalgaard}, \&
  {Wilcox}}]{Schatten1978}
{Schatten}, K.~H., {Scherrer}, P.~H., {Svalgaard}, L., \& {Wilcox}, J.~M. 1978,
  \grl, 5, 411, \dodoi{10.1029/GL005i005p00411}

\bibitem[{{Scherrer} {et~al.}(1995){Scherrer}, {Bogart}, {Bush}, {Hoeksema},
  {Kosovichev}, {Schou}, {Rosenberg}, {Springer}, {Tarbell}, {Title},
  {Wolfson}, {Zayer}, \& {MDI Engineering Team}}]{MDI}
{Scherrer}, P.~H., {Bogart}, R.~S., {Bush}, R.~I., {et~al.} 1995, \solphys,
  162, 129, \dodoi{10.1007/BF00733429}

\bibitem[{{Scherrer} {et~al.}(2012){Scherrer}, {Schou}, {Bush}, {Kosovichev},
  {Bogart}, {Hoeksema}, {Liu}, {Duvall}, {Zhao}, {Title}, {Schrijver},
  {Tarbell}, \& {Tomczyk}}]{HMI}
{Scherrer}, P.~H., {Schou}, J., {Bush}, R.~I., {et~al.} 2012, \solphys, 275,
  207, \dodoi{10.1007/s11207-011-9834-2}

\bibitem[{{Sheeley}(1966)}]{Sheeley1966}
{Sheeley}, N.~R., J. 1966, \apj, 144, 723, \dodoi{10.1086/148651}

\bibitem[{{Sreedevi} {et~al.}(2023){Sreedevi}, {Jha}, {Karak}, \&
  {Banerjee}}]{Sreedevi2023}
{Sreedevi}, A., {Jha}, B.~K., {Karak}, B.~B., \& {Banerjee}, D. 2023, arXiv
  e-prints, arXiv:2304.06615, \dodoi{10.48550/arXiv.2304.06615}

\bibitem[{{Virtanen} {et~al.}(2017){Virtanen}, {Virtanen}, {Pevtsov}, {Yeates},
  \& {Mursula}}]{Virtanen2017}
{Virtanen}, I.~O.~I., {Virtanen}, I.~I., {Pevtsov}, A.~A., {Yeates}, A., \&
  {Mursula}, K. 2017, \aap, 604, A8, \dodoi{10.1051/0004-6361/201730415}

\bibitem[{{Wang} \& {Sheeley}(1989)}]{WangYM1989}
{Wang}, Y.~M., \& {Sheeley}, N.~R., J. 1989, \solphys, 124, 81,
  \dodoi{10.1007/BF00146521}

\bibitem[{{Wang} \& {Sheeley}(1991)}]{Wang1991}
---. 1991, \apj, 375, 761, \dodoi{10.1086/170240}

\bibitem[{{Wang} \& {Sheeley}(2009)}]{Wang2009}
{Wang}, Y.~M., \& {Sheeley}, N.~R. 2009, \apjl, 694, L11,
  \dodoi{10.1088/0004-637X/694/1/L11}

\bibitem[{{Wang} {et~al.}(2021){Wang}, {Jiang}, \& {Wang}}]{Wang2021}
{Wang}, Z.-F., {Jiang}, J., \& {Wang}, J.-X. 2021, \aap, 650, A87,
  \dodoi{10.1051/0004-6361/202140407}

\bibitem[{{Whitbread} {et~al.}(2018){Whitbread}, {Yeates}, \&
  {Mu{\~n}oz-Jaramillo}}]{Whitbread2018}
{Whitbread}, T., {Yeates}, A.~R., \& {Mu{\~n}oz-Jaramillo}, A. 2018, \apj, 863,
  116, \dodoi{10.3847/1538-4357/aad17e}

\bibitem[{{Yeates}(2020)}]{2020Yeates}
{Yeates}, A.~R. 2020, \solphys, 295, 119, \dodoi{10.1007/s11207-020-01688-y}

\bibitem[{{Yeates} {et~al.}(2015){Yeates}, {Baker}, \& {van
  Driel-Gesztelyi}}]{Yeates2015}
{Yeates}, A.~R., {Baker}, D., \& {van Driel-Gesztelyi}, L. 2015, \solphys, 290,
  3189, \dodoi{10.1007/s11207-015-0660-9}

\bibitem[{{Yeates} {et~al.}(2023){Yeates}, {Cheung}, {Jiang}, {Petrovay}, \&
  {Wang}}]{Yeates2023}
{Yeates}, A.~R., {Cheung}, M. C.~M., {Jiang}, J., {Petrovay}, K., \& {Wang},
  Y.-M. 2023, \ssr, 219, 31, \dodoi{10.1007/s11214-023-00978-8}

\bibitem[{{Yeates} {et~al.}(2007){Yeates}, {Mackay}, \& {van
  Ballegooijen}}]{Yeates2007}
{Yeates}, A.~R., {Mackay}, D.~H., \& {van Ballegooijen}, A.~A. 2007, \solphys,
  245, 87, \dodoi{10.1007/s11207-007-9013-7}

\bibitem[{{Zhang} {et~al.}(2010){Zhang}, {Wang}, \& {Liu}}]{detectionZhang}
{Zhang}, J., {Wang}, Y., \& {Liu}, Y. 2010, \apj, 723, 1006,
  \dodoi{10.1088/0004-637X/723/2/1006}

\end{thebibliography}
\bibliographystyle{aasjournal}

\end{document}